\documentstyle[12pt,epsfig]{article}
\renewcommand{\baselinestretch}{2}
\begin{document}

\title{Electronic and optical properties in graphane\\}
\author{M. H. Lee$^{a}$,  H. C. Chung$^{a}$, J. M. Lu$^{c,d,*}$, C. P. Chang$^{b,*}$, and M. F. Lin$^{a,*}$ \\
\small $^a$Department of Physics, National Cheng Kung University,
 Tainan, Taiwan\\
\small $^b$Center for General Education, Tainan University of Technology,
 Tainan, Taiwan\\
\small $^c$National Center for High-Performance Computing,
 Tainan, Taiwan\\
\small $^d$Department of Mechanical Engineering, South Taiwan University,
 Tainan, Taiwan\\
\small{$^*$Email: 0403817@narlabs.org.tw (J.M. Lu)}\\
\small{$^*$Email: t00252@mail.tut.edu.tw (C.P. Chang)}\\
\small{$^*$Email: mflin@mail.ncku.edu.tw (M.F. Lin)}\\
}

\renewcommand{\baselinestretch}{1}
\date {}
\maketitle
\renewcommand{\baselinestretch}{1.5}
\begin{abstract}
We develop the tight-binding model to study electronic and optical properties of graphane.
The strong $sp^3$ chemical bondings among the carbon and hydrogen atoms induce
 a special band structure and thus lead to the rich optical excitations. The absorption spectrum hardly depends on the direction of electric polarization. It exhibits a lot of shoulder structures and absorption peaks, which arise from the extreme points and the saddle points of the parabolic bands, respectively. The threshold optical excitations, only associated with the $2p_x$ and $2p_y$ orbitals of the carbon atoms, are revealed in a shoulder structure at $\sim$3.5 eV.
The first symmetric absorption peak, appearing at $\sim$11 eV, corresponds to energy bands
due to the considerable hybridization of carbon $2p_z$ orbitals and H 1s orbitals. Also, some absorption peaks at higher frequencies indicate the bonding of $2s$ and $1s$ orbitals. These results are in sharp contrast to those of the $sp^2$ graphene systems.



\pagebreak
\end{abstract}
\renewcommand{\baselinestretch}{2}

\newpage

\noindent {\large{\bf 1. Introduction}}

Carbon-related materials contain zero- to three-dimensional symmetric structures, e.g., zero dimensional (0D) fullerenes \cite{1}, 0D carbon tori \cite{2}, 1D carbon nanotubes \cite{3,4,5,6}, 1D nanographene ribbons \cite{7,8,9,10,11}, 2D ultrathin graphite films \cite{12,13,14,15}, and
3D layered graphites \cite{16,17}. Since the 2D Graphene, a one-atom-thick planar sheet of carbon atoms, was successfully produced in 2004 \cite{12}, many theoretical and experimental studies have focused on its essential physical properties, e.g., electronic properties \cite{18,19,20,21,22,23}, transport properties \cite{13,24,25,26,27}, optical spectra \cite{18,19,28,29,30,31,36,37}, and Coulomb excitations \cite{32,33,34,35}. This system is a zero-gap semiconductor with a crossing point of two linear $\pi$ bands at the Fermi level, and is potentially an ideal candidate for  next-generation electronic devices due to the performance of a high electron mobility. However, the application of graphene is limited because of the absence of an energy gap. In order to further open up a sufficiently large energy gap, many methods have been applied to modulate the electronic properties of graphene, e.g., electric field \cite{19,36,37,38}, magnetic field \cite{27,39}, mechanical strain \cite{40,41,42}, and chemical adsorption \cite{43,44,45}. Especially, doping with other kinds of atoms in graphene alters the intrinsic properties drastically, and the resulting material exhibits many different physical properties. In this research, we focus on how the adsorption of hydrogen atoms completely change the electronic and optical properties.

A new hydrocarbon material called graphane was predicted by Sofo et al. in 2007 \cite{46}, utilizing the first-principles density functional theory (DFT) method \cite{47,48,49}. They reported that the added hydrogen atoms change the carbon atoms from $sp^{2}$ hybridization to $sp^{3}$ hybridization, and the band gap of the stable chair-like conformation is 3.5 eV. Subsequently, graphane was investigated in other theoretical works using the DFT method. Also, possible applications of graphane in electronic devices \cite{50,51,52} have been studied theoretically. On the experimental side, Elias et al. \cite{53} successfully manufactured graphane in 2009. They hydrogenated graphene via exposing it to a cold hydrogen plasma, and the predicted gap-opening phenomenon was verified by the experimental  measurements of the electric conductance. Furthermore, Balog et al \cite{54}. controlled the hydrogen adsorption by exposing graphene to hydrogen for different lengths of time, in which the hydrogen density and the energy gap grow with increasing exposure time. Apparently, the electronic properties are dramatically changed by the density of adsorpted hydrogen atoms.

The tight-binding model\cite{4,6} is developed to calculate the electronic and optical properties of graphane. The energy dispersions have the critical points (extreme and saddle points) in the energy-wave-vector space at the highly symmetric points, e.g. the $\Gamma$ , M, and K points; therefore, the energy bands induce special structures (shoulder structures and symmetric peaks) in the density of states (DOS). Furthermore, the DOS could be decomposed into the contributions from different orbitals, then the effects due to the chemical bondings and the main features of the optical excitations could be explained. The predicted DOS could be verified by scanning tunneling spectroscopy (STS) \cite{55,56}. The main characteristics of the energy bands are reflected in the optical absorption spectra and evaluated by the Fermi golden rule and the gradient approximation\cite{5,6,8,9,10,18,19,57}. The mechanisms  of optical excitations and the dependence on the direction of the electric polarization are investigated in detail. This work shows that graphane exhibits the rich optical spectra, which include many shoulder structures and prominent peaks, and a high threshold absorption frequency. Those features could be observed by utilizing experimental optical spectroscopy measurements\cite{28,29,30,36,37}. In comparison to graphene, the electronic and optical properties of graphane exhibit very different features; these will be explored.

\vskip 0.6 truecm
\noindent {\large{\bf 2. The tight-binding model theory}}

The two dimensional structure of the graphane in the chair-like configuration is shown in Fig. 1. Two carbon atoms in a unit cell are connected with two hydrogen atoms in opposite directions. Carbon-carbon bondings are formed by the $sp^{3}$ hybridization of four orbitals (2s, 2$p_{x}$, $2p_{y}$, $2p_{z}$), and carbon-hydrogen bondings are associated with the hybridization of $sp^{3}$ and 1s orbital. The C-C and C-H bond lengths, represented as $b$ and $b^{\prime}$, are about 1.52 {\AA } and 1.1 {\AA }, respectively. Note that the bonding angles of C-C-H and C-C-C are nearly identical at about $\theta=109.5^{\circ}$, owing to the tetrahedral arrangement. The first Brillouin zone (1st BZ) is defined in a hexagonal reciprocal lattice (Fig. 1). The vectors of the three highly symmetric points are  $\Gamma$(0,0), M($\frac{2\pi}{\sqrt{3}b^{\prime\prime}}$,0), and K($\frac{2\pi}{\sqrt{3}b^{\prime\prime}}$,$\frac{2\pi}{3b^{\prime\prime}}$), where $b^{\prime\prime}$ is 2$b\sin(\frac{\theta}{2})$. The electronic structure of graphane is calculated with the tight-binding model including the on-site and nearest-neighbor atomic interactions. Such interactions include the $\pi$ bondings, $\sigma$ bondings (Fig. 2; subscripts in the following V's) and on-site energies. The tight-binding parameters are approximated by $V_{2s,2p\sigma}$=$-$4.22 eV, $V_{2s,2s\sigma}$=$-$5.10 eV, $V_{2p,2p\sigma}$=$-$4.20 eV,  $V_{1s,2p\sigma}$=$-$5.58 eV, $V_{1s,2s\sigma}$=$-$3.65 eV, $V_{2p,2p\pi}$=$-$2.38 eV, $\xi_{2s}$=$-$4.8 eV, $\xi_{2p}$=$-$2.5 eV; $\xi_{1s}$=$-$4.0 eV. It should also be noted that the atomic interactions vanish for the 1s and 2$p_{x}$ ($2p_{y}$) orbitals. The Bloch wave function arising from the linear superposition of the ten tight-binding functions is expressed as:
\begin{eqnarray}
 \begin{array}{l}
\left| {\mathop \Psi \nolimits^{c,v} } \right\rangle =a_1\left| {2s^A} \right\rangle + a_2\left| {2p_x^A} \right\rangle  + a_3\left| {2p_y^A} \right\rangle + a_4\left| {2p_z^A} \right\rangle + a_5\left| {1s^A}
\right\rangle \\
 \quad\quad\quad +b_1\left| {2s^B} \right\rangle + b_2\left| {2p_x^B} \right\rangle + b_3\left| {2p_y^B} \right\rangle + b_4\left| {2p_z^B} \right\rangle
 + b_5\left| {1s^B} \right\rangle,
\end{array}
\end{eqnarray}
where $|2p_{x,y}^{A,B}\rangle$ ($|2s^{A,B}\rangle$) is the tight-binding function of the $2p_{x,y}$ (2s) orbitals from the carbon atoms, and $|1s^{A,B}\rangle$ is the tight-binding function of the 1s orbital from the hydrogen atoms. The superscript A(B) denotes the carbon or hydrogen atom at an A(B) site in the unit cell. The $10\times10$ Hermitian Hamiltonian matrix is built from the subspaces in the sequence of
$\{|2s^{A}\rangle,|2p_{x}^{A}\rangle,|2p_{y}^{A}\rangle,|2p_{z}^{A}\rangle,|1s^{A}\rangle,|2s^{B}\rangle,|2p_{x}^{B}\rangle,|2p_{y}^{B}\rangle,|2p_{z}^{B}\rangle;|1s^{B}\rangle\}$.
Therefore, the non-zero independent matrix elements are given by
\begin{eqnarray}
\begin{array}{l}
H_{11}=\langle 2s^{A}|\textbf{H}|2s^{A}\rangle=\xi_{2s},
\end{array}
\end{eqnarray}

\begin{eqnarray}
\begin{array}{l}
H_{15}=\langle 2s^{A}|\textbf{H}|1s^{A}\rangle=V_{1s,2s\sigma},
\end{array}
\end{eqnarray}

\begin{eqnarray}
\begin{array}{l}
H_{22}=\langle 2p_{x}^{A}|\textbf{H}|2p_{x}^{A}\rangle=\xi_{2p},
\end{array}
\end{eqnarray}

\begin{eqnarray}
\begin{array}{l}
H_{33}=\langle 2p_{y}^{A}|\textbf{H}|2p_{y}^{A}\rangle=\xi_{2p},
\end{array}
\end{eqnarray}

\begin{eqnarray}
\begin{array}{l}
H_{44}=\langle 2p_{z}^{A}|\textbf{H}|2p_{z}^{A}\rangle=\xi_{2p},
\end{array}
\end{eqnarray}

\begin{eqnarray}
\begin{array}{l}
H_{45}=\langle 2p_{z}^{A}|\textbf{H}|1s^{A}\rangle=V_{1s,2p\sigma},
\end{array}
\end{eqnarray}

\begin{eqnarray}
\begin{array}{l}
H_{55}=\langle 1s^{A}|\textbf{H}|1s^{A}\rangle=\xi_{1s},
\end{array}
\end{eqnarray}

\begin{eqnarray}
\begin{array}{l}
H_{66}=\langle 2s^{B}|\textbf{H}|2s^{B}\rangle=\xi_{2s},
\end{array}
\end{eqnarray}

\begin{eqnarray}
\begin{array}{l}
H_{6,10}=\langle 2s^{B}|\textbf{H}|1s^{B}\rangle=V_{1s,2s\sigma},
\end{array}
\end{eqnarray}

\begin{eqnarray}
\begin{array}{l}
H_{77}=\langle 2p_{x}^{B}|\textbf{H}|2p_{x}^{B}\rangle=\xi_{2p},
\end{array}
\end{eqnarray}

\begin{eqnarray}
\begin{array}{l}
H_{88}=\langle 2p_{y}^{B}|\textbf{H}|2p_{y}^{B}\rangle=\xi_{2p},
\end{array}
\end{eqnarray}

\begin{eqnarray}
\begin{array}{l}
H_{99}=\langle 2p_{z}^{B}|\textbf{H}|2p_{z}^{B}\rangle=\xi_{2p},
\end{array}
\end{eqnarray}

\begin{eqnarray}
\begin{array}{l}
H_{9,10}=\langle 2p_{z}^{B}|\textbf{H}|1s^{B}\rangle=V_{1s,2p\sigma},
\end{array}
\end{eqnarray}

\begin{eqnarray}
\begin{array}{l}
H_{10,10}=\langle 1s^{B}|\textbf{H}|1s^{B}\rangle=\xi_{1s},
\end{array}
\end{eqnarray}

\begin{eqnarray}
\begin{array}{l}
H_{16}=\langle 2s^{A}|\textbf{H}|2s^{B}\rangle\\=\exp(ik_{x}\frac{b^{\prime\prime}}{\sqrt{3}})V_{2s,2s\sigma}+2\exp(-ik_{x}\frac{b^{\prime\prime}}{2\sqrt{3}})V_{2s,2s\sigma}\cos(\frac{b^{\prime\prime}}{2}k_{y}),
\end{array}
\end{eqnarray}

\begin{eqnarray}
\begin{array}{l}
H_{17}=\langle 2s^{A}|\textbf{H}|2p_{x}^{B}\rangle \\=\exp(-ik_{x}\frac{b^{\prime\prime}}{2\sqrt{3}}-ik_{y}\frac{b^{\prime\prime}}{2})V_{2s,2p\sigma}\cos\theta_{3x}
\\
-\exp(ik_{x}\frac{b^{\prime\prime}}{\sqrt{3}})V_{2s,2p\sigma}\cos\theta_{1x}+\exp(-ik_{x}\frac{b^{\prime\prime}}{2\sqrt{3}}+ik_{y}\frac{b^{\prime\prime}}{2})V_{2s,2p\sigma}\cos\theta_{2x},
\end{array}
\end{eqnarray}

\begin{eqnarray}
\begin{array}{l}
H_{18}
=\langle 2s^{A}|\textbf{H}|2p_{y}^{B}\rangle\\=-\exp(-ik_{x}\frac{b^{\prime\prime}}{2\sqrt{3}}+ik_{y}\frac{b^{\prime\prime}}{2})V_{2s,2p\sigma}\cos\theta_{2y}+\exp(-ik_{x}\frac{b^{\prime\prime}}{2\sqrt{3}}-ik_{y}\frac{b^{\prime\prime}}{2})V_{2s,2p\sigma}\cos\theta_{3y},
\end{array}
\end{eqnarray}

\begin{eqnarray}
\begin{array}{l}
H_{19}=\langle 2s^{A}|\textbf{H}|2p_{z}^{B}\rangle
\\
=\exp(ik_{x}\frac{b^{\prime\prime}}{\sqrt{3}})V_{2s,2p\sigma}\cos\theta_{1z}+\exp(-ik_{x}\frac{b^{\prime\prime}}{2\sqrt{3}}+ik_{y}\frac{b^{\prime\prime}}{2})V_{2s,2p\sigma}\cos\theta_{2z}
\\
+\exp(-ik_{x}\frac{b^{\prime\prime}}{2\sqrt{3}}+ik_{y}\frac{b^{\prime\prime}}{2})V_{2s,2p\sigma}\cos\theta_{3z},
\end{array}
\end{eqnarray}

\begin{eqnarray}
\begin{array}{l}
H_{26}=\langle 2p_{x}^{A}|\textbf{H}|2s^{B}\rangle
\\
=\exp(ik_{x}\frac{b^{\prime\prime}}{\sqrt{3}})V_{2s,2p\sigma}\cos\theta_{1x}-\exp(-ik_{x}\frac{b^{\prime\prime}}{2\sqrt{3}}+ik_{y}\frac{b^{\prime\prime}}{2})V_{2s,2p\sigma}\cos\theta_{2x}
\\
-\exp(-ik_{x}\frac{b^{\prime\prime}}{2\sqrt{3}}-ik_{y}\frac{b^{\prime\prime}}{2})V_{2s,2p\sigma}\cos\theta_{3x},
\end{array}
\end{eqnarray}

\begin{eqnarray}
\begin{array}{l}
H_{27}=\langle 2p_{x}^{A}|\textbf{H}|2p_{x}^{B}\rangle\\=\exp(ik_{x}\frac{b^{\prime\prime}}{\sqrt{3}})\{-V_{2p,2p\sigma}(\cos\theta_{1x})^{2}+V_{2p,2p\pi}(\sin\theta_{1x})^{2}\}
\\
+\exp(-ik_{x}\frac{b^{\prime\prime}}{2\sqrt{3}}+ik_{y}\frac{b^{\prime\prime}}{2})\{-V_{2p,2p\sigma}(\cos\theta_{2x})^{2}+V_{2p,2p\pi}(\sin\theta_{2x})^{2}\}
\\
+\exp(-ik_{x}\frac{b^{\prime\prime}}{2\sqrt{3}}-ik_{y}\frac{b^{\prime\prime}}{2})\{-V_{2p,2p\sigma}(\cos\theta_{3x})^{2}+V_{2p,2p\pi}(\sin\theta_{3x})^{2}\},
\end{array}
\end{eqnarray}

\begin{eqnarray}
\begin{array}{l}
H_{28}=\langle 2p_{x}^{A}|\textbf{H}|2p_{y}^{B}\rangle\\=\exp(ik_{x}\frac{b^{\prime\prime}}{\sqrt{3}})\{V_{2p,2p\sigma}\cos\theta_{1x}\cos\theta_{1y}
+V_{2p,2p\pi}\sin\theta_{1x}\sin\theta_{1y}\cos\alpha_{1xy}\}
\\
+\exp(-ik_{x}\frac{b^{\prime\prime}}{2\sqrt{3}}+ik_{y}\frac{b^{\prime\prime}}{2})\{V_{2p,2p\sigma}\cos\theta_{2x}\cos\theta_{2y}
+V_{2p,2p\pi}\sin\theta_{2x}\sin\theta_{2y}\cos\alpha_{2xy}\}
\\
+\exp(-ik_{x}\frac{b^{\prime\prime}}{2\sqrt{3}}-ik_{y}\frac{b^{\prime\prime}}{2})\{-V_{2p,2p\sigma}\cos\theta_{3x}\cos\theta_{3y}
+V_{2p,2p\pi}\sin\theta_{3x}\sin\theta_{3y}\cos\alpha_{3xy}\},
\end{array}
\end{eqnarray}

\begin{eqnarray}
\begin{array}{l}
H_{29}=\langle 2p_{x}^{A}|\textbf{H}|2p_{z}^{B}\rangle\\=\exp(ik_{x}\frac{b^{\prime\prime}}{\sqrt{3}})\{V_{2p,2p\sigma}\cos\theta_{1x}\cos\theta_{1z}
+V_{2p,2p\pi}\sin\theta_{1x}\sin\theta_{1z}\cos\alpha_{1xz}\}
\\
+\exp(-ik_{x}\frac{b^{\prime\prime}}{2\sqrt{3}}+ik_{y}\frac{b^{\prime\prime}}{2})\{-V_{2p,2p\sigma}\cos\theta_{2x}\cos\theta_{2z}
+V_{2p,2p\pi}\sin\theta_{2x}\sin\theta_{2z}\cos\alpha_{2xz}\}
\\
+\exp(-ik_{x}\frac{b^{\prime\prime}}{2\sqrt{3}}-ik_{y}\frac{b^{\prime\prime}}{2})\{-V_{2p,2p\sigma}\cos\theta_{3x}\cos\theta_{3z}
+V_{2p,2p\pi}\sin\theta_{3x}\sin\theta_{3z}\cos\alpha_{3xz}\} ,
\end{array}
\end{eqnarray}

\begin{eqnarray}
\begin{array}{l}
H_{36}=\langle 2p_{y}^{A}|\textbf{H}|2s^{B}\rangle\\=2i\exp(-ik_{x}\frac{b^{\prime\prime}}{2\sqrt{3}})V_{2s,2p\sigma}\sin(\frac{w}{2}k_{y})\cos\theta_{2y},
\end{array}
\end{eqnarray}

\begin{eqnarray}
\begin{array}{l}
H_{37}=\langle 2p_{y}^{A}|\textbf{H}|2p_{x}^{B}\rangle\\=\exp(ik_{x}\frac{b^{\prime\prime}}{\sqrt{3}})\{V_{2p,2p\sigma}\cos\theta_{1y}\cos\theta_{1x}
+V_{2p,2p\pi}\sin\theta_{1y}\sin\theta_{1x}\cos\alpha_{1xy}\}
\\
+\exp(-ik_{x}\frac{b^{\prime\prime}}{2\sqrt{3}}+ik_{y}\frac{b^{\prime\prime}}{2})\{V_{2p,2p\sigma}\cos\theta_{2x}\cos\theta_{2y}
+V_{2p,2p\pi}\sin\theta_{2x}\sin\theta_{2y}\cos\alpha_{2xy}\}
\\
+\exp(-ik_{x}\frac{b^{\prime\prime}}{2\sqrt{3}}-ik_{y}\frac{b^{\prime\prime}}{2})\{-V_{2p,2p\sigma}\cos\theta_{3x}\cos\theta_{3y}
+V_{2p,2p\pi}\sin\theta_{3x}\sin\theta_{3y}\cos\alpha_{3xy}\},
\end{array}
\end{eqnarray}

\begin{eqnarray}
\begin{array}{l}
H_{38}=\langle 2p_{y}^{A}|\textbf{H}|2p_{y}^{B}\rangle\\=\exp(-ik_{x}\frac{b^{\prime\prime}}{2\sqrt{3}})\{V_{2p,2p\sigma}(\cos\theta_{1y})^{2}+V_{2p,2p\pi}(\sin\theta_{1y})^{2}\} \\
+\exp(-ik_{x}\frac{b^{\prime\prime}}{2\sqrt{3}}+ik_{y}\frac{b^{\prime\prime}}{2})\{-V_{2p,2p\sigma}(\cos\theta_{2y})^{2}+V_{2p,2p\pi}(\sin\theta_{2y})^{2}\} \\
+\exp(-ik_{x}\frac{b^{\prime\prime}}{2\sqrt{3}}-ik_{y}\frac{b^{\prime\prime}}{2})\{-V_{2p,2p\sigma}(\cos\theta_{3y})^{2}+V_{2p,2p\pi}(\sin\theta_{3y})^{2}\},
\end{array}
\end{eqnarray}

\begin{eqnarray}
\begin{array}{l}
H_{39}=\langle 2p_{y}^{A}|\textbf{H}|2p_{z}^{B}\rangle\\=\exp(ik_{x}\frac{b^{\prime\prime}}{\sqrt{3}})\{V_{2p,2p\sigma}\cos\theta_{1y}\cos\theta_{1z}
+V_{2p,2p\pi}\sin\theta_{1y}\sin\theta_{1z}\cos\alpha_{1yz}\}
\\
+\exp(-ik_{x}\frac{b^{\prime\prime}}{2\sqrt{3}}+ik_{y}\frac{b^{\prime\prime}}{2})\{V_{2p,2p\sigma}\cos\theta_{2y}\cos\theta_{2z}
+V_{2p,2p\pi}\sin\theta_{2y}\sin\theta_{2z}\cos\alpha_{2yz}\}
\\
+\exp(-ik_{x}\frac{b^{\prime\prime}}{2\sqrt{3}}-ik_{y}\frac{b^{\prime\prime}}{2})\{-V_{2p,2p\sigma}\cos\theta_{3z}\cos\theta_{3y}
+V_{2p,2p\pi}\sin\theta_{3z}\sin\theta_{3y}\cos\alpha_{3yz}\},
\end{array}
\end{eqnarray}

\begin{eqnarray}
\begin{array}{l}
H_{46}=\langle 2p_{z}^{A}|\textbf{H}|2s^{B}\rangle
\\=-\exp(ik_{x}\frac{b^{\prime\prime}}{\sqrt{3}})V_{2s,2p\sigma}\cos\theta_{1z}-2\exp(-ik_{x}\frac{b^{\prime\prime}}{2\sqrt{3}})V_{2s,2p\sigma}\cos(\frac{b^{\prime\prime}}{2}k_{y})\cos\theta_{2z},
\end{array}
\end{eqnarray}

\begin{eqnarray}
\begin{array}{l}
H_{47}=\langle 2p_{z}^{A}|\textbf{H}|2p_{x}^{B}\rangle\\=\exp(ik_{x}\frac{b^{\prime\prime}}{\sqrt{3}})\{V_{2p,2p\sigma}\cos\theta_{1x}\cos\theta_{1z}
+V_{2p,2p\pi}\sin\theta_{1x}\sin\theta_{1z}\cos\alpha_{1xz}\}
\\
+\exp(-ik_{x}\frac{b^{\prime\prime}}{2\sqrt{3}}+ik_{y}\frac{b^{\prime\prime}}{2})\{-V_{2p,2p\sigma}\cos\theta_{2x}\cos\theta_{2z}
+V_{2p,2p\pi}\sin\theta_{2x}\sin\theta_{2z}\cos\alpha_{2xz}\}
\\
+\exp(-ik_{x}\frac{b^{\prime\prime}}{2\sqrt{3}}-ik_{y}\frac{b^{\prime\prime}}{2})\{-V_{2p,2p\sigma}\cos\theta_{3z}\cos\theta_{3x}
+V_{2p,2p\pi}\sin\theta_{3z}\sin\theta_{3x}\cos\alpha_{3xz}\} ,
\end{array}
\end{eqnarray}

\begin{eqnarray}
\begin{array}{l}
H_{48}=\langle 2p_{z}^{A}|\textbf{H}|2p_{y}^{B}\rangle\\=\exp(ik_{x}\frac{b^{\prime\prime}}{\sqrt{3}})\{V_{2p,2p\sigma}\cos\theta_{1y}\cos\theta_{1z}
+V_{2p,2p\pi}\sin\theta_{1y}\sin\theta_{1z}\cos\alpha_{1yz}\}
\\
+\exp(-ik_{x}\frac{b^{\prime\prime}}{2\sqrt{3}}+ik_{y}\frac{b^{\prime\prime}}{2})\{-V_{2p,2p\sigma}\cos\theta_{2y}\cos\theta_{2z}
+V_{2p,2p\pi}\sin\theta_{2y}\sin\theta_{2z}\cos\alpha_{2yz}\}
\\
+\exp(-ik_{x}\frac{b^{\prime\prime}}{2\sqrt{3}}-ik_{y}\frac{b^{\prime\prime}}{2})\{-V_{2p,2p\sigma}\cos\theta_{3z}\cos\theta_{3y}
+V_{2p,2p\pi}\sin\theta_{3z}\sin\theta_{3y}\cos\alpha_{3yz}\},
\end{array}
\end{eqnarray}

\begin{eqnarray}
\begin{array}{l}
H_{49}=\langle 2p_{z}^{A}|\textbf{H}|2p_{z}^{B}\rangle\\=\exp(ik_{x}\frac{b^{\prime\prime}}{\sqrt{3}})\{-V_{2p,2p\sigma}(\cos\theta_{1z})^{2}+V_{2p,2p\pi}(\sin\theta_{1z})^{2}\} \\
+\exp(-ik_{x}\frac{b^{\prime\prime}}{2\sqrt{3}}+ik_{y}\frac{b^{\prime\prime}}{2})\{-V_{2p,2p\sigma}(\cos\theta_{2z})^{2}+V_{2p,2p\pi}(\sin\theta_{2z})^{2}\} \\
+\exp(-ik_{x}\frac{b^{\prime\prime}}{2\sqrt{3}}-ik_{y}\frac{b^{\prime\prime}}{2})\{-V_{2p,2p\sigma}(\cos\theta_{3z})^{2}+V_{2p,2p\pi}(\sin\theta_{3z})^{2}\},
\end{array}
\end{eqnarray}
$\theta_{1x}$ is the angle defined between the unit vector $\hat{x}$ and the position vector $\overrightarrow{R}_{1}$. $\theta_{2x}$ and $\theta_{3x}$ are the angles with reference to the position vectors $\overrightarrow{R}_{2}$ and $\overrightarrow{R}_{3}$ (Fig. 1). Similarly to $\theta_{1x}$, the angles $\theta_{1y}$, $\theta_{2y}$, $\theta_{3y}$, $\theta_{1z}$, $\theta_{2z}$, and $\theta_{3z}$ are defined.
On the other hand, the $2p_{x}$ and $2p_{y}$ orbitals are projected to the position vectors $\overrightarrow{R}_{1}$ ($\overrightarrow{R}_{2}$,$\overrightarrow{R}_{3}$) with their normal component vectors $2p_{xn}$, and $2p_{yn}$ respectively. Then, $\alpha_{1xy}$ ($\alpha_{2xy}$, $\alpha_{3xy}$) is the twist angle between the two normal component vectors $2p_{xn}$ and $2p_{yn}$. Similarly, the other twist angels $\alpha_{1yz}$, $\alpha_{2yz}$, $\alpha_{3yz}$, $\alpha_{1xz}$, $\alpha_{2xz}$, and $\alpha_{3xz}$ are defined. By solving the Hamiltonian matrix, the energy dispersion $E^{c,v}$ and the wave function $\psi^{c,v}$ are obtained.

\vskip 0.6 truecm
\noindent  {\large{\bf 3. Electronic and optical properties}}

The $sp^{3}$ bonding graphane exhibits a multitude of electronic properties, which are quite different from those of the $sp^{2}$ bonding graphene. In graphane, the energy bands of occupied states and unoccupied states are not symmetric about the Fermi energy $E_{F}=0$, as shown in Fig. 3 along the highly symmetric points $K\rightarrow \Gamma \rightarrow M \rightarrow K$ in the 1st BZ (Fig. 1). The conduction and valence bands away from $E_{F}=0$ can conveniently be characterized with the band indices $n^{c}$ and $n^{v}$. The energy bands have parabolic dispersions and many band-edge states. Such energy dispersions induce the special structures in the density of states (DOS) at critical points which are either the extreme value points or saddle points. Most of the extreme points are located at the $\Gamma$ or K point, while the saddle points are all located at the M point. Meanwhile, the energy gap $E_{g}=3.52$ eV between the highest valence state and the lowest conduction state is observed at the $\Gamma$ point. Furthermore, the smallest energy difference at the M point is determined by the $n^{c}=1$ conduction band and the $n^{v}=1$ valence band as $E^{s}_{M}\simeq 11$ eV. At the K point, the energy difference is $E^{s}_{K}\simeq 13$ eV. Compared with graphane, the electronic properties of graphene display very different characteristics. Graphene possesses two distinct groups of energy bands ($\pi,\pi^{\ast}$) and ($\sigma, \sigma^{\ast}$). The former, originating from the the $2p_{z}$ orbitals, are $\pi$ bands, which are degenerate at the K point with the Dirac cone; in consequence, the energy gap vanishes. The latter, associated with the hybridization of 2$s$, $2p_{x}$ and $2p_{y}$ orbitals, are $\sigma$ bands about 2 eV away from the Fermi energy.

The DOS, which reveals the main characteristics of the two dimensional band structures, is defined as
\begin{equation}
D(\omega)=\sum_{c,v}{\int_{1st BZ}}{{dk_xdk_y} \over
(2\pi)^2}\frac{\Gamma}{\pi[({\omega}-E^{c,v}(k_x,k_y) )^{2}+ {\Gamma}^2]}
\end{equation}
where $\Gamma=$10 meV is the phenomenological broadening parameter. The total DOS (black curve in Fig. 4) is composed of two features. One is the symmetric peaks in the logarithmically divergent form, which come from the band structures at the saddle point M. The other is the shoulder structures induced by the band-edge states at extreme-value points  ($\Gamma$ and K points). Furthermore, the total DOS can be considered as the combination of the DOS from the 2$s$, 1$s$, $2p_{x}+2p_{y}$, and $2p_{z}$ orbitals (colored curves in Fig. 4). The DOS from each orbital is useful in understanding the electronic structures, and the main characteristics of optical spectra. The low-lying electronic states mainly arise from the $2p_{x}$ and $2p_{y}$ orbitals, since the atomic interactions almost vanish between these orbitals and the 1$s$ orbital.
Such orbitals create two shoulder structures in the DOS at $\omega=\pm$ 1.76 eV; these are associated with the band-edge states at the $\Gamma$ point. They also make important contributions to the other special structures over a wide frequency range. On the other hand, the strong hybridization of $2p_z$ and 1$s$ orbitals leads to the initial contributions at $\omega=-$ 4.0 eV and 3.0 eV, where two shoulder structures are presented. The succeeding contribution from the hybridization of these two orbitals displays the  prominent symmetric peaks associated with the saddle point M at $\omega=-$ 6 eV ($\omega=$ 5 eV) for valence (conduction) states. The other important contributions for higher energies also exhibit the special structures in DOS between $-$11 eV$<\omega<$$-$6 eV (5 eV$<\omega<$11 eV) for the valence (conduction) states. In comparison with graphene, the strong hybridization in graphane causes the energy bands dominated by $2p_z$ orbital to fall in the valence states and to rise in the conduction states. Unlike the $2p_{x}$ and $2p_{y}$ orbitals, the 2$s$ orbital primarily contributes to the deeper  (higher)  valence (conduction) bands. Particularly, the lowest valence and the highest conduction states are only contributed by the 2$s$ orbital. Also, the two shoulder structures in DOS corresponding to those two states are created at $\Gamma$ point. It should be noted that the highest DOS peak contributed by all  the orbitals is revealed at $\omega$=$-$8 eV ($\omega$=6.5 eV) for valence (conduction) states. The special DOS structures in the $\pm$6 eV range could be verified by scanning tunneling spectroscopy (STS) \cite{55,56}, an experiment being successfully performed on carbon-related materials. The tunneling electrical conductance is proportional to the DOS and reflects the special structures in DOS.

At zero temperature, the electrons of graphane are vertically exited from occupied valence states to unoccupied conduction states when an electromagnetic field with a polarization of $\hat{\mathbf{E}}//\hat{x}$ or $\hat{\mathbf{E}}//\hat{y}$ is applied; this means that the initial and the final states have the same wave vector. According to Fermi's golden rule, the optical absorption function $A(\omega)$ from the vertical transition is given by:
\begin{eqnarray}
\begin{array}{l}
 A(\omega ) \propto \sum\limits_{c,v,n^{c},n^{v}} {\int_{1stBZ} {\frac{{dk_{x}dk_{y} }}{{(2\pi)^{2} }}} }  \times {\mathop{\rm Im}\nolimits} \left[ {\frac{{f[E^{c} (k_{x},k_{y},n^{c})] - f[E^v (k_{x},k_{y},n^{v})]}}{{E^{c} (k_{x},k_{y},n^{c}) - E^v (k_{x},k_{y},n^{v}) - \omega  - i\Gamma }}} \right] \\
 \quad \quad \quad \quad \quad \quad \;\;\,\quad \quad \quad \quad \quad  \times \left| {\left\langle {\psi ^{c} (k_{x},k_{y},n^{c})\left| {\frac{{{\rm{\hat{\textbf{E}}}} \cdot \mathord{\buildrel{\lower3pt\hbox{$\scriptscriptstyle\rightharpoonup$}}
\over P} }}{{m_e }}} \right|\psi ^v (k_{x},k_{y},n^{v})} \right\rangle } \right|^2 , \\
 \end{array}
\end{eqnarray}
where $f[E^{v} (k_{x},k_{y},n^{c})]$ ($f[E^c (k_{x},k_{y},n^{v})]$) is the Fermi-Dirac distribution of the valence (conduction) states, and $\Gamma$=10 meV is the phenomenological broadening parameter. $A(\omega)$ is determined by the joint density of states (JDOS), and the velocity matrix element: $\left| {\left\langle {\psi ^{c} (k_{x},k_{y},n^{c})\left| {\frac{{{\rm{\hat{\textbf{E}}}} \cdot \mathord{\buildrel{\lower3pt\hbox{$\scriptscriptstyle\rightharpoonup$}}
\over P} }}{{m_e }}} \right|\psi ^v (k_{x},k_{y},n^{v})} \right\rangle } \right| $. The former represents the available channels in the vertical optical excitations. The square of the latter represents the intensity of each excitation channel, and could be evaluated within the gradient approximation. Such approximation has been successfully used to explain optical spectra of carbon-related systems, such as carbon nanotubes \cite{5,6}, graphite \cite{57}, and few-layer graphenes \cite{18,19}. Graphane exhibits rich optical absorption spectra, mainly owing to the strong chemical bonding between the hydrogen and carbon.
The JDOS, as shown in Fig. 5(a), contains two kinds of special structures, i.e., shoulder structures and symmetric peaks in the logarithmically divergent form. Such features are associated with the vertical optical excitations arising from the highly symmetric points ($\Gamma$, M, and K); they are directly reflected in $A(\omega)$ (Fig. 5(b)).

$A(\omega)$ is not sensitive to a change in the direction of the electric polarization. The dependence of $A(\omega)$ on the frequency is similar to that of JDOS; that is, the main features of the optical spectra are dominated by the number of excitation channels. The first shoulder structure of $A(\omega)$ near $\omega$=3.52 eV comes from the optical transition between $E^{v}$(-1.76 eV)$\rightarrow$ $E^{c}$(1.76 eV) for $n^{v}=$1 $\rightarrow$ $n^{c}=$1 at the band-edge states associated with the $\Gamma$ point. The second, third, and fourth shoulder structures, respectively, represent the optical transitions $E^{v}$(-1.76 eV)$\rightarrow$ $E^{c}$(3 eV) for $n^{v}=$1$\rightarrow$ $n^{c}=$3, $E^{v}$(-4.0 eV) $\rightarrow$ $E^{c}$(1.76 eV) for $n^{v}=$3 $\rightarrow$ $n^{c}=$1, and $E^{v}$(-4.0 eV) $\rightarrow$ $ E^{c}$(3 eV) for $n^{v}=$3 $\rightarrow$ $n^{c}=$3 at the $\Gamma$ point. The strength of the first shoulder, dominated by $2p_x$ and $2p_y$ orbitals, is the weakest one. The fourth shoulder structure is stronger than the others because the strong hybridization of the $2p_{z}$ and 1$s$ orbitals starts to make important contributions to optical excitations. The other shoulders also exist at higher frequency ($\omega>$22 eV), and the highest-frequency shoulder induced only by the 2$s$ orbitals is as weak as the first one.

The absorption spectrum also shows other symmetric peaks in the frequency range of 11$-$22 eV, owing to the vertical optical transitions between the conduction and valence states at the saddle point M. The strong DOS at the M point causes the stronger intensity for the symmetric peaks than for the shoulders. The first symmetric absorption peak appearing at $\omega\simeq$11 eV (blue circle) comes from the transition $E^{v}$(-6 eV)$\rightarrow$ $E^{c}$(5 eV) for $n^{v}=$1 $\rightarrow$ $n^{c}=$1; this transition is induced by the strong hybridization of the $2p_z$ and 1s orbitals. Unlike the $sp^{3}$ graphane, $sp^{2}$ carbon-related systems including graphite, graphene, graphene nanoribbons, and carbon nanotube exhibit the prominent absorption peaks at ${\omega\sim\,4-6}$ eV, which is the most important feature of $\pi$-electronic optical excitations. This directly reflects the fact that the low-lying $\pi$ band of the $2p_z$ orbitals in the $sp^2$ systems change into the deep-lying hybridized band of $2p_z$ and 1s orbitals in the $sp^3$ systems. The strongest one among all the absorption peaks occurs at $\omega\simeq$13 eV (green circle) and is associated with the optical transitions $E^{v}$(-6.5 eV)$\rightarrow$ $E^{c}$(6.5 eV) for $n^{v}=$2 $\rightarrow$ $n^{c}=$2, and $E^{v}$(-8 eV)$\rightarrow$ $E^{c}$(5 eV) for $n^{v}=$3 $\rightarrow$ $n^{c}=$1 at the M point, $E^{v}$(-10 eV)$\rightarrow$ $E^{c}$(3 eV) for $n^{v}=$4 $\rightarrow$ $n^{c}=$3 at the $\Gamma$ point, and $E^{v}$(-7 eV)$\rightarrow$ $E^{c}$(6 eV) for $n^{v}=$1 $\rightarrow$ $n^{c}=$1 at the K point. Specifically, the absorption peak at $\omega\simeq$14.5 eV (pink circle) originates from $n^{v}=$3 $\rightarrow$ $n^{c}=$2 at the M point, and represents the transition between the two states which posses the strongest DOS owing to the contributions from all orbitals at $\omega=-$8 eV for the valence states, and $\omega=$6.5 eV for the conduction states. Other absorption peaks associated with the optical transitions at the highly symmetric points in the frequency range 11$-$22 eV also exist. Parts of those absorption peaks are overlaps of two or more transitions since the frequency of some transitions are located close to each other.

Four special features exist in the absorption spectrum: the threshold absorption frequency only coming from the $2p_{x}$ and $2p_{y}$ orbitals, the first symmetric peak associated with the strong hybridization of the $2p_{z}$ and $1s$ orbitals,
the strongest peak originating from four available optical transitions, and the absorption peak at $\omega\simeq$14.5 eV which is contributed by all orbitals. Optical spectroscopy can be used to verify these important features and the strong chemical bondings underlying them. In comparison with the optical absorption spectra of graphene, the large blue shift of the $2p_z$-dependent absorption peak related to the strong hybridization can be observed by utilizing experimental optical spectroscopy measurements\cite{28,29,30,36,37}.

\vskip 0.6 truecm
\noindent  {\large{\bf 4. Conclusion}}

The tight-binding model is developed to study the electronic and optical properties of graphane associated with its strong $sp^{3}$ hybridization. Many band-edge states in parabolic dispersions exist. Most of the extreme points are located at the $\Gamma$ or K point, whereas the saddle points are all located at the M point. Meanwhile, the direct energy gap at the $\Gamma$ point is $E_{g}=3.52$ eV. The DOS displays shoulder structures and symmetric peaks in  logarithmically divergent form for the energy bands at the highly symmetric points. From the orbital-dependent DOS, the low-lying electronics are identified to mainly arise from the $2p_{x}$ and $2p_{y}$ orbitals. The symmetric prominent peaks at $\omega=-$ 6 eV ($\omega=$ 5 eV) for the valence (conduction) states indicate that the strong hybridization of $2p_{z}$ and $1s$ orbitals changes the low-lying $\pi$ bands in $sp^{2}$ graphene into the deep-lying hybridized bands in the $sp^3$ systems. Also, the 2$s$ orbital primarily contributes to the deeper  (higher)  valence (conduction) bands. The predicted special structures of DOS in the $\pm$6 eV range could be examined by STS measurements; they are reflected in the tunneling electrical conductance.

Graphane exhibits rich optical absorption spectra, which are dominated by the joint density of states, but not by the electric dipole momenta. The main features of optical properties are insensitive to a change in the direction of the electric polarization. The threshold optical excitations, which are only induced by the $2p_{x}$ and $2p_{y}$ orbitals, reveal the shoulder structures at $\omega\simeq$3.52 eV. The first symmetric absorption peak due to the strong hybridization of $2p_{z}$ and $1s$ orbitals occurs at $\omega\simeq$11 eV. Also, some absorption peaks at higher frequencies indicate the chemical bonding of 2s and 1s orbitals. On the other hand, the threshold frequency for $sp^{2}$ graphene systems is almost zero, and the $\pi$-electronic optical excitations display the prominent absorption peaks at ${\omega\sim\,4-6}$ eV. In graphane, a large blue shift of the the $2p_z$-dependent absorption peak is apparent. The main features of the absorption spectra could be verified by the experimental optical spectroscopy measurements.

\vskip 0.6 truecm
\noindent  {\large{\bf Acknowledgment}}

This work was supported in part by the Ministry of Science and Technology of
Taiwan, the Republic of China under Grant No. NSC 102-2112-M-006-007-MY3,
No. NSC 102-2112-M-165-001-MY3, and No. MOST 103-2218-E-038-002.

\newpage
{\Large\bf Figure captions}
\renewcommand{\baselinestretch}{1}
\begin{itemize}

\item[Fig. 1:](Color online) (a) Side view and
(b) top view of graphane, (c) the first Brillouin zone with symmetry points $\Gamma$, M, and K; (d) the position vectors from the A atom to the neighboring B atoms.

\item[Fig. 2:](Color online) Vanishing and non-vanishing orbital interactions.

\item[Fig. 3:](Color online) Energy dispersions of graphane.

\item[Fig. 4:](Color online) The total density of states (DOS), and the projected DOSs for $2s$, $1s$, $2p_{x}+2p_{y}$, and $2p_{z}$ orbitals.

\item[Fig. 5:](Color online). (a) The joint density of states , and (b) the absorption spectrum for two directions of electric polarization ($\hat{\mathbf{E}}//\hat{x}$ and $\hat{\mathbf{E}}//\hat{y}$).

\end{itemize}

\end{document}